\documentclass[twocolumn,showpacs,preprintnumbers,amsmath,amssymb]{revtex4}
\usepackage{tabularx,graphicx}\begin{document}
\newcommand{\beq}{\begin{equation}}
\newcommand{\eeq}{\end{equation}}
\newcommand{\beqn}{\begin{eqnarray}}
\newcommand{\eeqn}{\end{eqnarray}}
\newcommand{\bmath}{\begin{subequations}}
\newcommand{\emath}{\end{subequations}}
\newcommand{\bra}[1]{\langle #1|}
\newcommand{\ket}[1]{|#1\rangle}

\title{Spherical agglomeration of superconducting and normal microparticles with and without applied electric field}
\author{R.S.B. Ghosh and J. E. Hirsch}
\address{Department of Physics, University of California, San Diego,
La Jolla, CA 92093-0319}

\begin{abstract} 
It was   reported  by R. Tao and coworkers that in the presence of a strong electric field superconducting microparticles assemble
into  balls of macroscopic dimensions. Such a finding has potentially important implications for the understanding of the fundamental physics of
superconductors. However, we report here the results of experimental studies showing  that (i) ball formation also occurs in the absence of
an applied electric field,    (ii)  the phenomenon    also occurs at temperatures  above the superconducting transition 
temperature, and (iii) it can also  occur for non-superconducting materials.  Possible origins of the phenomenon   are   discussed.

\end{abstract}
\maketitle 

\section{introduction} 
It has been reported by R. Tao and coworkers that superconducting microparticles in a strong electric field assemble
into compact balls  of macroscopic dimensions (diameter fraction of a mm) \cite{tao,tao2,tao3,tao4}.  These balls  are robust and survive collisions with the electrodes at high speeds. 
Tao et al attributed the ball formation  to a new kind of ``surface tension'' of superconducting
particles in the presence of a strong electric field \cite{tao}. They found  that this phenomenon occurs for both high $T_c$ cuprate superconductors\cite{tao}
and for low $T_c$ conventional superconductors\cite{tao2,tao4}. Furthermore they reported that the balls become fragile and break apart easily both when the
electric field is turned off and/or when the temperature is raised above the superconducting transition temperature.

If  valid, these findings would have enormous  implications for the understanding of the fundamental physics of superconductors. The conventional theory of superconductivity\cite{tinkham} predicts 
that the response of superconductors to applied electrostatic fields is the same as that of normal metals\cite{bs}: electric fields should be  screened within a distance on the order of Angstroms
(Thomas-Fermi length) of the surface, and thus would not be expected\cite{tao}  to give rise to phenomena  such as the spherical aggregation
reported by Tao. One of us has proposed\cite{taoeffect}  that the phenomenon reported by Tao
 is evidence of the inadequacy of the conventional theory of superconductivity\cite{bcsvalidity}, and is 
a consequence of the same physics that leads to the Meissner effect within the unconventional theory of
``hole superconductivity''\cite{holesc}, namely expulsion of negative charge from the interior of the superconductor to the surface. Quite generally, if this aggregation indeed took place only
in the superconducting state and only under a strong applied electric field, it would call for an explanation beyond the generally accepted physics of superconductors.

 However, we report in this paper that the  spherical aggregation  is a more widespread phenomenon than previously reported. We find that
  it can occur both in the presence and in the absence of an electric field, and both  in the  superconducting state as well as above the
  superconducting transition temperature, and also in non-superconducting materials. These findings undermine the claims that the Tao experiments cast doubt on the conventional theory of
  superconductivity\cite{bcsvalidity} and are evidence in favor of an alternative theory\cite{taoeffect}. While the phenomenon appears to be strongest in materials that are high temperature
  superconductors, we have not found clear evidence that it is stronger in the superconducting versus the nonsuperconducting state.
  
 \section{agglomeration in an electric field} 
 
 We started by attempting to reproduce Tao's experimental results. In addition to Tao's papers, one unpublished report exists that independently supported some of
  Tao's findings\cite{taoconfirm}.

 We used
$BSCCO$ and  $YBa_2Cu_3O_{7-\delta}$   disks  provided by Colorado Superconductor. The $BSCCO$ disks were reportedly a mixture of the 2223 and the 2212 phases, with critical temperature above $100 K$. Both types of disks exhibited a strong Meissner effect at liquid nitrogen
temperatures.
We ground the disks with a Dremel (952 aluminum oxide grinding stone) and sorted particles by sizes using sieves. Figure 1 shows typical shapes of particles obtained for various sizes.
We used only the resulting microparticles that 
passed through a $32 \mu m$ sieve.   In another set of experiments, we used $Bi_2Sr_2CaCu_2O_{8+x}$,  $x=0.15-0.20)$ 
powder provided by
Sigma-Aldrich, hereafter termed `$\sigma-$powder', with nominal grain size $\leq 5 \mu m$.
 
    \begin{figure}
 \resizebox{7.5cm}{!}{\includegraphics[width=8cm]{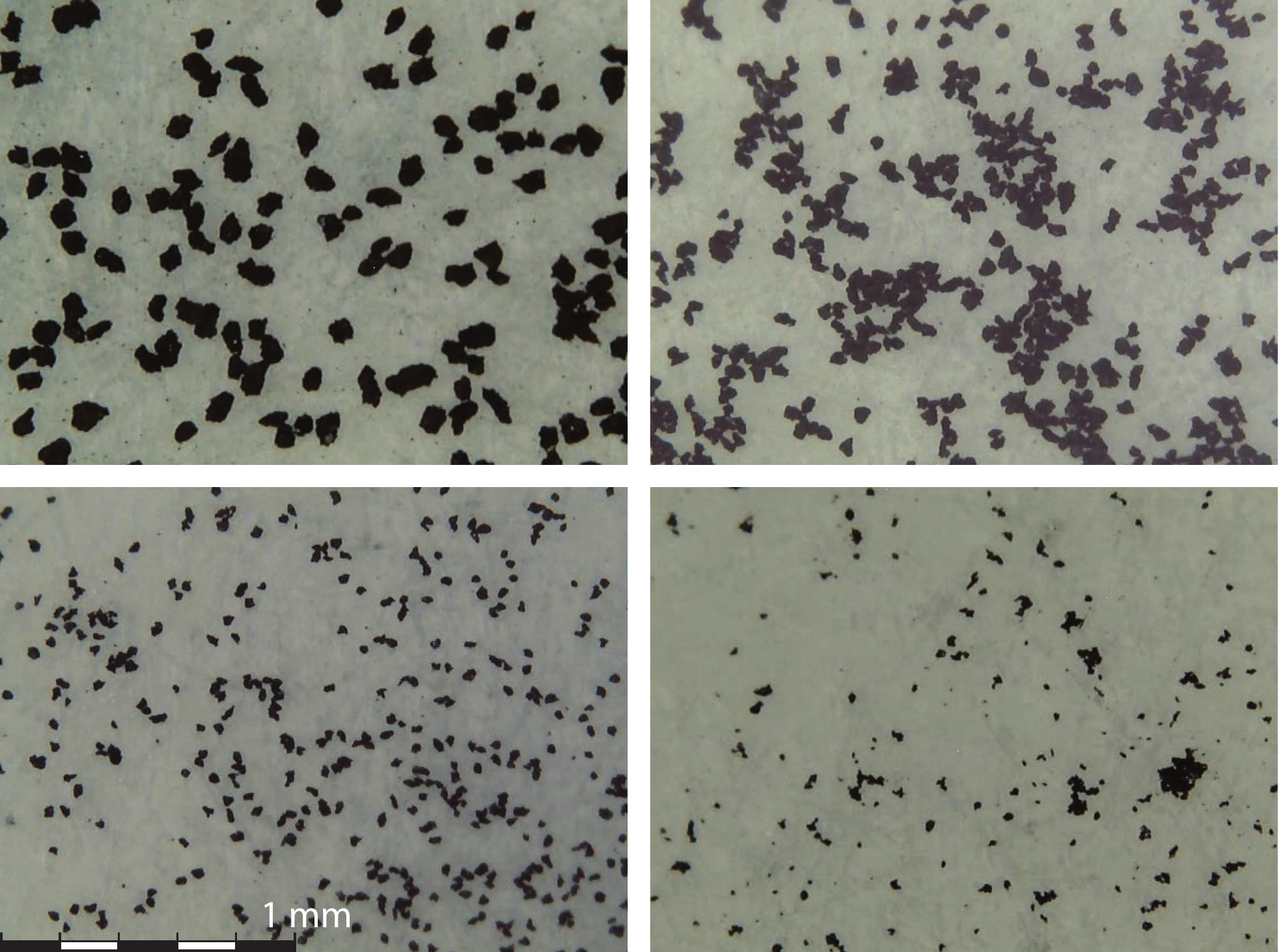}}
 \caption {Typical microparticles of various sizes obtained by grinding superconducting disks and passing through sieves. The range of diameters of the four panels are $53\mu m$ to $75\mu m$, 
 $32\mu m$ to $53\mu m$, $25\mu m$ to $32\mu m$,  and less than $32 \mu m$ for the upper left, upper right, lower left and lower right panels.
 Note the irregular shapes of the particles.}
 \label{chain}
 \end{figure}
  
 We constructed a cell measuring $15mm \times 70mm$ horizontally and $13mm$ vertically   from plastic plates of thickness $1 mm$, and attached thin ($<1mm$) metal plates  either on the inside or outside walls. This formed a parallel plate capacitor with electrodes separated by $12mm$ to $16mm$ depending on the configuration. We used both $Al$ and stainless steel metal plates and observed the same behavior,
 indicating that magnetism was not playing a role. The metal plates  were connected to a DC voltage source that could deliver up to 20kV, i.e. maximum electric field
 inside the capacitor approximately $1700 V/mm$. This is well above the `lower critical field'   $E_{c1}$ required for ball formation reported by Tao of order $500-800 V/mm$\cite{tao,tao2,tao3,tao4}.
 
  We poured liquid nitrogen into the capacitor,
added $BSCCO$ or $YBCO$ powder (typically between $60 mg$ and $150 mg$) and observed the behavior when the voltage source was connected to the capacitor plates by
recording  with a high speed camera that yielded 1000 frames per second.

When we put both metal plates on the inside of the plastic plates we got interesting action starting at a few thousand volts, as described below. When both metal plates were outside
of the plastic plates to form insulating electrodes we found no action whatsoever, even for voltages up to 20kV. When one of the metal plates was on the inside and one on the outside we found
some very limited action. We conclude that the effect is $not$ driven by a large electric field, as postulated by Tao\cite{tao,tao2,tao3,tao4}, 
since it  requires $charge$ being able to
circulate from the conducting electrodes to the superconducting microparticles. 

   \begin{figure}
 \resizebox{8.5cm}{!}{\includegraphics[width=9cm]{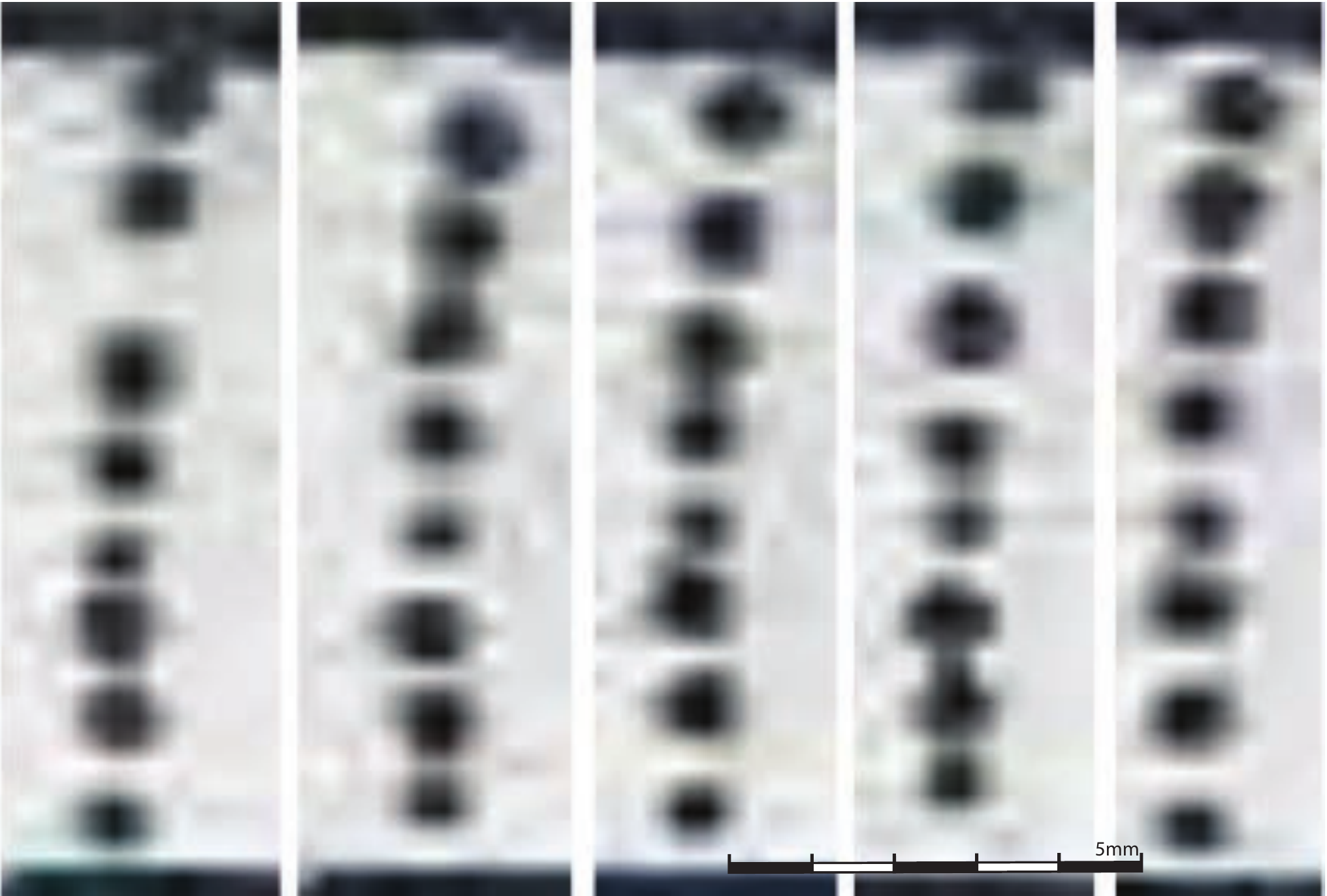}}
 \caption { Balls bouncing up and down between electrodes, frames shown are consecutive separated by 10 ms. The vertical size of the figure corresponds
 to 1cm in real life. Each ball is composed of several thousand micrograins}
 \label{chain}
 \end{figure}

The action we found in the capacitor can be summarized as follows. On increasing the voltage, at several thousand volts, powder starts to move. Sometimes it suddenly flies to the electrodes. Then, balls form and start bouncing between the electrodes, as described by Tao. In addition, as described by Tao, in all runs clinging of powder to the electrodes occurs. The voltage where these processes occurs is variable, depending for example on the amount of powder and its distribution between the electrodes, it was typically in the range between 6kV and 14kV.  While Tao usually found only one
ball in his $4mm \times 5mm$ chamber, in our setup we typically found a large number of balls of varying size. 
Often, the balls alligned in a linear fashion and bounced back and forth against each other and against the electrodes.  Figure \ref{chain} shows as an
example a row of 8  balls in five snapshots separated by 10 frames each, i.e 10 ms (the camera records 1000 frames per second),
in a run where the potential difference between top and bottom electrodes was 9kV.
The distance between top and bottom electrodes (black horizontal lines) is 1 cm, so these balls range from approximately 0.45 mm to 0.70 mm in diameter. 
Note that a ball of 0.5 mm diameter would be composed of at least 4,000   microparticles  in close proximity.
The balls
are bouncing up and down and there is no apparent deformation when they bounce.  Here they move between collisions at a speed of approximately $6 cm/s$,
 and  the acceleration involved in a collision is
of order  $10 m/s^2$. This linear arrangement was stable for approximately 2s, corresponding to each particle traveling the equivalent of 12 times
the distance between the electrodes and undergoing about 200 collisions with the neighboring particles. At the end of that period a ninth particle squeezed into the row and the same arrangement continued until shedding from the layer
coating the upper electrode occurred.

The fact that ball formation only occurred when the electrodes were conducting indicates that the phenomenon is not due to the presence of an electric field
as argued by Tao, rather it appears
to be $facilitated$ by the  presence of   free charge. We hypothesize two processes: in the first process, 
microparticles pick up positive and negative charges from the electrodes, after which they are repelled from the electrodes and
return to the bulk of the chamber. Microparticles of opposite sign attract each other due to the long-range  Coulomb force. When they come in contact, they share their charge. One might 
expect that they would then separate, the fact that they stay together indicates that   another force  sets in that keeps them together.
In the second process, ball formation occurs at the electrodes themselves, where a thick layer of micrograins accumulates, with micrograins in close proximity to each other;
occasionally, pieces of the layer detach themselves in the form of balls that return to the bulk of the chamber. Again, the balls are held together by a force that acts
between micrograins in close proximity.

  \section{agglomeration in the absence of an electric field} 
\begin{figure}
 \resizebox{8.5cm}{!}{\includegraphics[width=9cm]{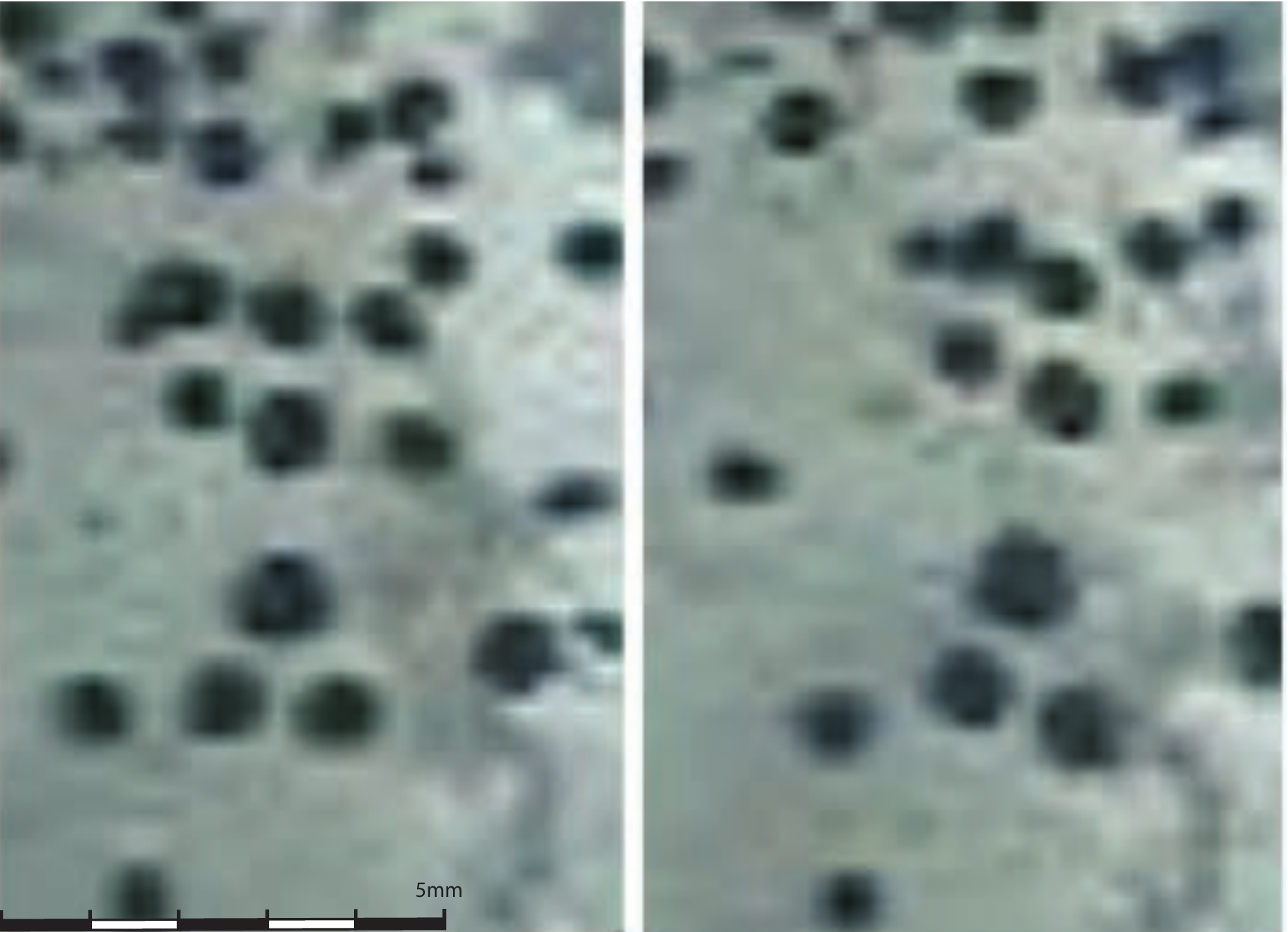}}
 \caption {$BSCCO$ balls formed by shaking a container with micrograins, no electric field applied. The scale is the same
 as in Figure 1. The two images are separated by 30 ms and the motion is due to motion of the nitrogen caused
 by shaking.}
 \label{nofield}
 \end{figure}

   We next   tested the possibility of sphere formation in the absence of applied  electric field and electric charges, 
   by gently shaking a container with
superconducting powder horizontally in liquid $N_2$.  
    We poured 60 mg of $BSCCO$ or $YBCO$ powder in a container of diameter  $80mm$ containing approximately $5mm$ height of $N_2$, and
   kept the container cold by submerging it in a container of larger diameter with substantially higher levels of $N_2$. In all cases the micrograins in the powders
   had diameter smaller than $32 \mu m$. Indeed, it was found after fine-tuning the procedure that it was very easy to make balls of 
the same sizes as produced in the capacitor. Figure \ref{nofield} shows one example. After formed, the balls retain their integrity upon strong  shaking of the container. When the agitation became sufficiently rapid and erratic the spheres would break apart back into powder. The balls could be reformed by returning to regular shaking.  In addition, we
verified that placing these balls inside the capacitor and applying a voltage between the plates leads to the same motion as described earlier for the case where
balls were created in the capacitor when voltage was applied.

   \section{temperature and material dependence} 
   Having established that ball formation in high temperature superconductors is independent of the presence or absence of an applied electric field, 
   we proceeded to investigate the temperature dependence of the effect and to test the behavior of various different materials. Tao et al had reported that both in high temperature superconductors,
   in low temperature elemental superconductors, and in magnesium diboride ($MgB_2$), balls don't form above $T_c$ and that balls that formed below 
   $T_c$ become fragile and are easily destroyed as soon as the temperature rises above $T_c$.
   In the following, we will present results for various materials in liquid $N_2$ (temperature $77.4K$) and liquid $Ar$ (temperature $87.3K$), as well as in
   water at room temperature and in ethanol and methanol at temperatures in the range  $150-180 K$. 
   
   \subsection{Materials used}
   We used the following materials:
   
   (1) $\sigma-$powder was pressed into a pellet of density approximately $3.5 g/cm^3$. It was heated to $850 ^oC$ and kept at that temperature in 
   air for $40$ hours, then cooled slowly to room temperature in another 40 hours. After this process the pellet did not exhibit a Meissner effect either in $N_2$ or in $Ar$. In addition,
   its transition temperature  was measured resistively and found to have onset at $T=73^oK$, confirming the observed absence of Meissner effects. The pellet was then ground and the resulting powder passed through a $32 \mu m$ sieve. The resulting powder collected in a dry tube and gently shaken had
   density $1.6 g /cm^3$.

(2) A $YBCO$ disk from Colorado Superconductors was broken into 3 pieces. One piece of the disk  was left intact.
This piece exhibited a strong Meissner effect in $N_2$ and a weaker Meissner effect in $Ar$. 
Another piece   was brought to   $800^oC$ for 90 minutes and then quenched to $77K$ in liquid $N_2$. According to the literature \cite{schuller,hariharan},
this procedure removes oxygen and leads to an underdoped sample, with $\delta \sim 0.65$ and $T_c$ less than $30K$. We did not measure the $T_c$ of this
sample but verified that it exhibited no Meissner effect in $N_2$. The sample was conducting at room temperature. 
We ground and sieved these samples the same way as described for sample (1).
The third piece was heated to $1050 ^oC$ for $120$ minutes and then cooled suddenly to $77K$. The resulting sample was now insulating, indicating
that more oxygen had been removed by this process, as expected, and of course exhibited no Meisser effect. Also this sample turned to a green color. We ground this sample the usual way, which turned out
to be considerably more difficult than in the other cases as this sample was considerably harder.

(3) $MgB_2$ powder obtained from
Alfa Aesar, with transition temperature expected to be $39K$. Again the powder was passed through a $32\mu m$ sieve.

(4) $Pb$ powder, ground from a bulk piece and passed through a $32\mu m$ sieve.

(5) $Zn$ powder passed through a $32\mu m$ sieve.

   \subsection{Results}

We found that ball formation occurs both below and above $T_c$, and that balls that formed below $T_c$ do not become fragile above $T_c$.
   Indeed, balls formed at low temperature remain robust even when they are brought to �room temperature.
 
\begin{figure}
 \resizebox{8.5cm}{!}{\includegraphics[width=9cm]{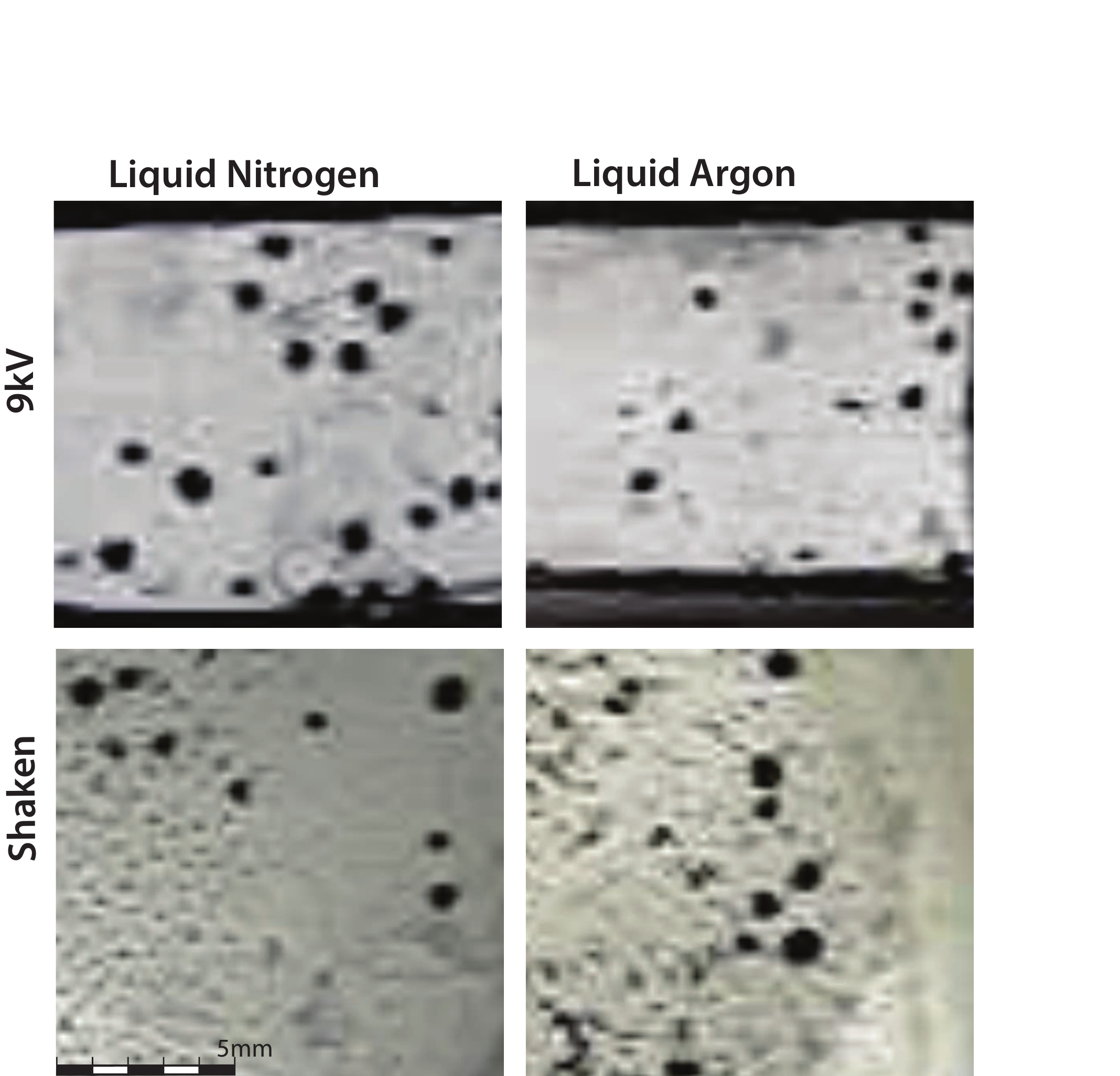}}
 \caption { $BSCCO$ $\sigma-$powder annealed at $850^oC$ inside the capacitor with voltage 9kV (upper panels) and obtained by shaking in the absence
 of electric field (lower panels). }
 \label{sigma4}
 \end{figure}

 Figure  \ref{sigma4} shows ball formation for the annealed $\sigma$-powder (see (1) in previous subsection) in $N_2$ and $Ar$, both in the capacitor with an electric field
 (upper panels)  and obtained
 by shaking without an electric field (lower panels). Since the critical temperature for this 
 sample was $73K$, all cases correspond to temperatures higher than $T_c$. It can be seen that the balls look very similar in all cases. 
 
\begin{figure}
 \resizebox{8.5cm}{!}{\includegraphics[width=9cm]{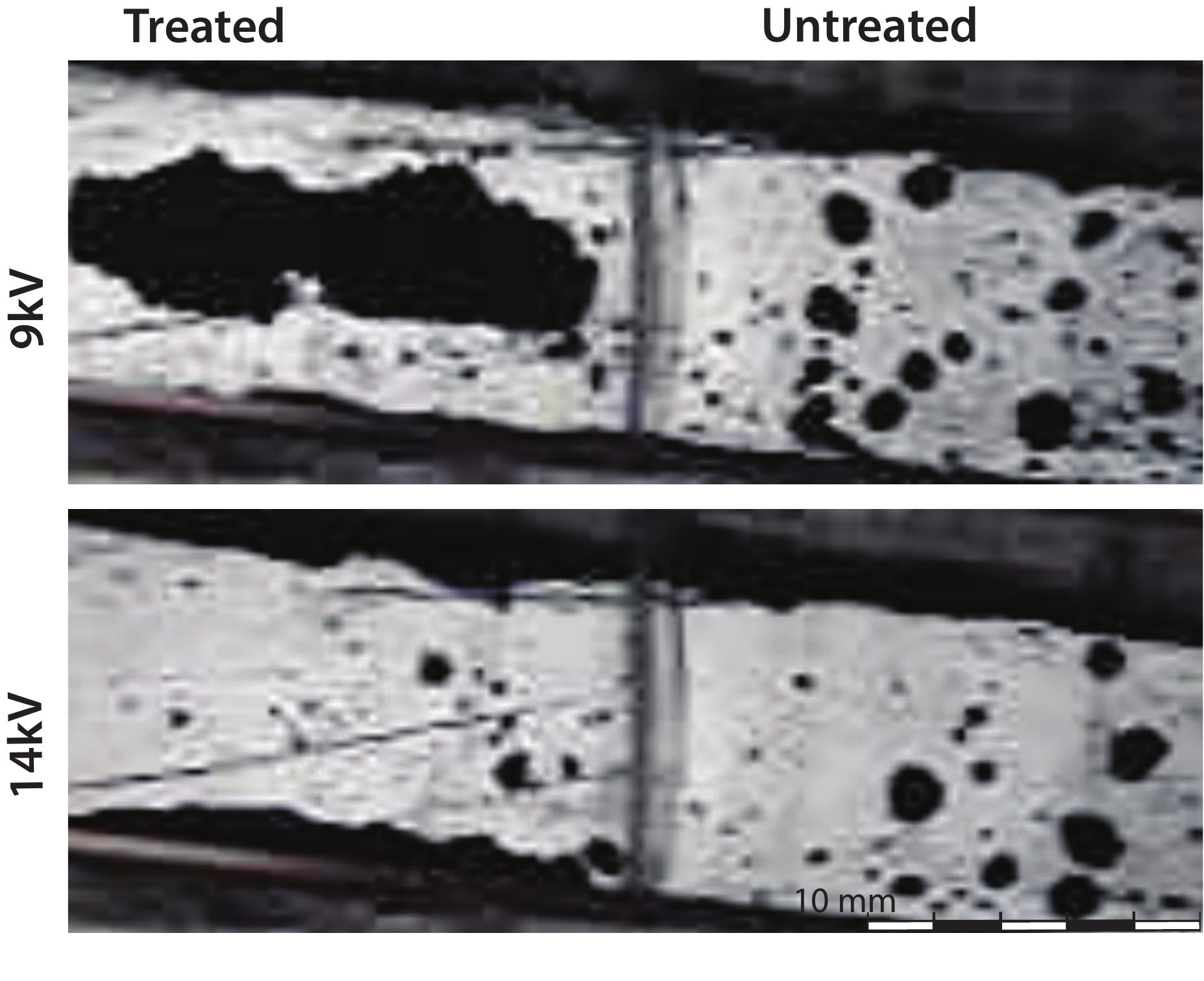}}
 \caption {$YBCO$ heat treated  to $800^oC$  (left panels) and untreated (right panels)  in capacitor  in liquid nitrogen.
 Voltage between capacitor plates is $9kV$ in upper panels and $14 kV$ in lower panels.}
 \label{ybcocooked}
 \end{figure}

Figure \ref{ybcocooked}  shows results for $YBCO$ heat treated  to $800^oC$ (left panels) (see (2) in previous subsection for description of heat treatment)
and untreated (right panels) 
 in the capacitor in $N_2$. Initially (upper panels) we brought the voltage between capacitor plates to
9kV. The untreated powder (which exhibits a Meissner effect) forms balls under this voltage, while the heat treated powder did not. When we then raised the voltage
to 14kV, the heated powder also formed balls  as seen in figure \ref{ybcocooked}.
There was also powder clinging to the electrodes both below and above $T_c$, as the right and left sides of figure   \ref{ybcocooked} illustrate. 

\begin{figure}
 \resizebox{8.5cm}{!}{\includegraphics[width=9cm]{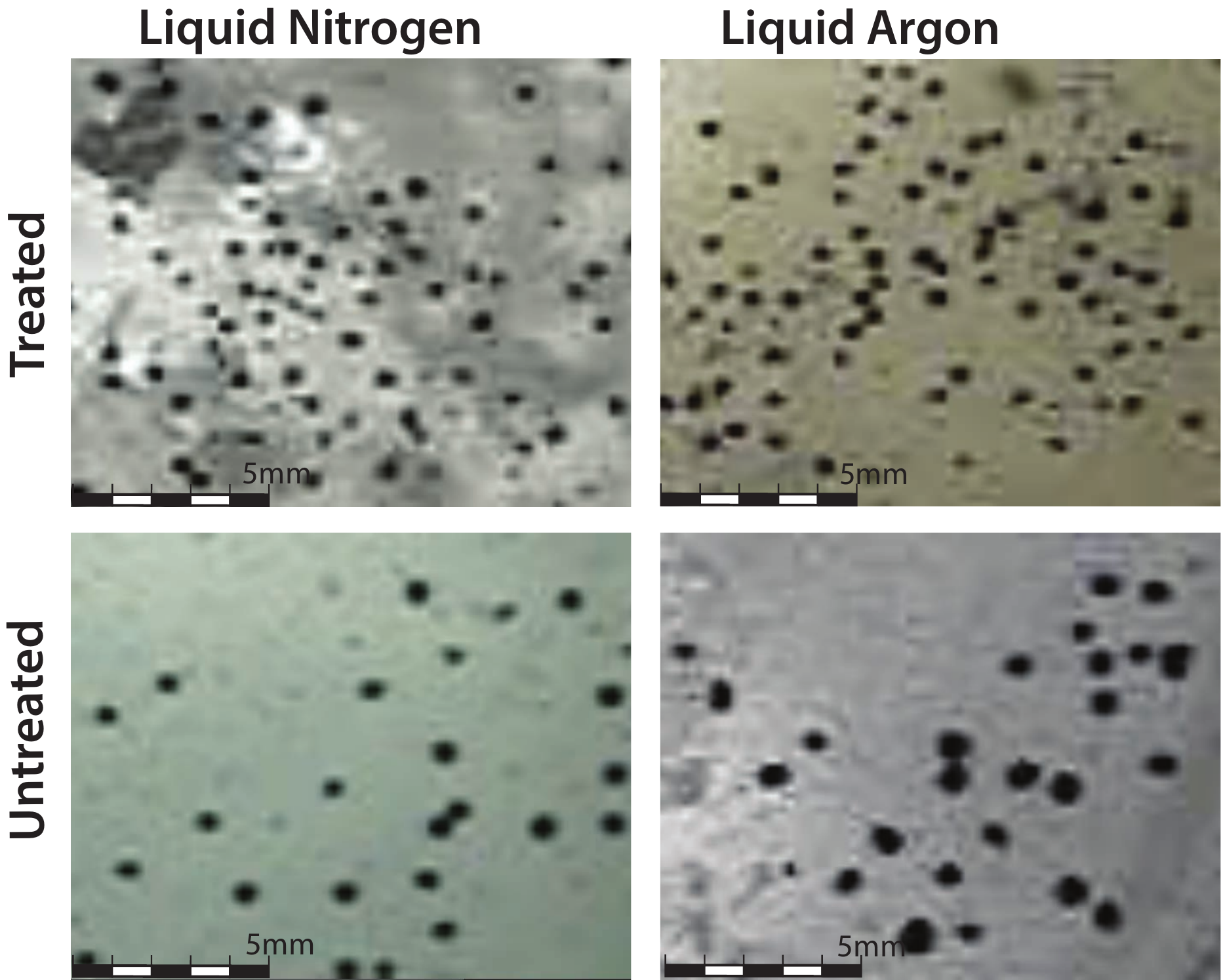}}
 \caption {$YBCO$ balls obtained by shaking in container with nitrogen and argon, no electric field. Upper panels: sample heat treateded to $800^oC$; lower panels: 
 untreated sample.}
 \label{ybcocooked2}
 \end{figure}

We also investigated ball formation under shaking in these samples, both in $N_2$ and $Ar$, and show representative images in Figure \ref{ybcocooked2}. Again no discernable difference
is seen among the four different cases.

\begin{figure}
 \resizebox{8.5cm}{!}{\includegraphics[width=9cm]{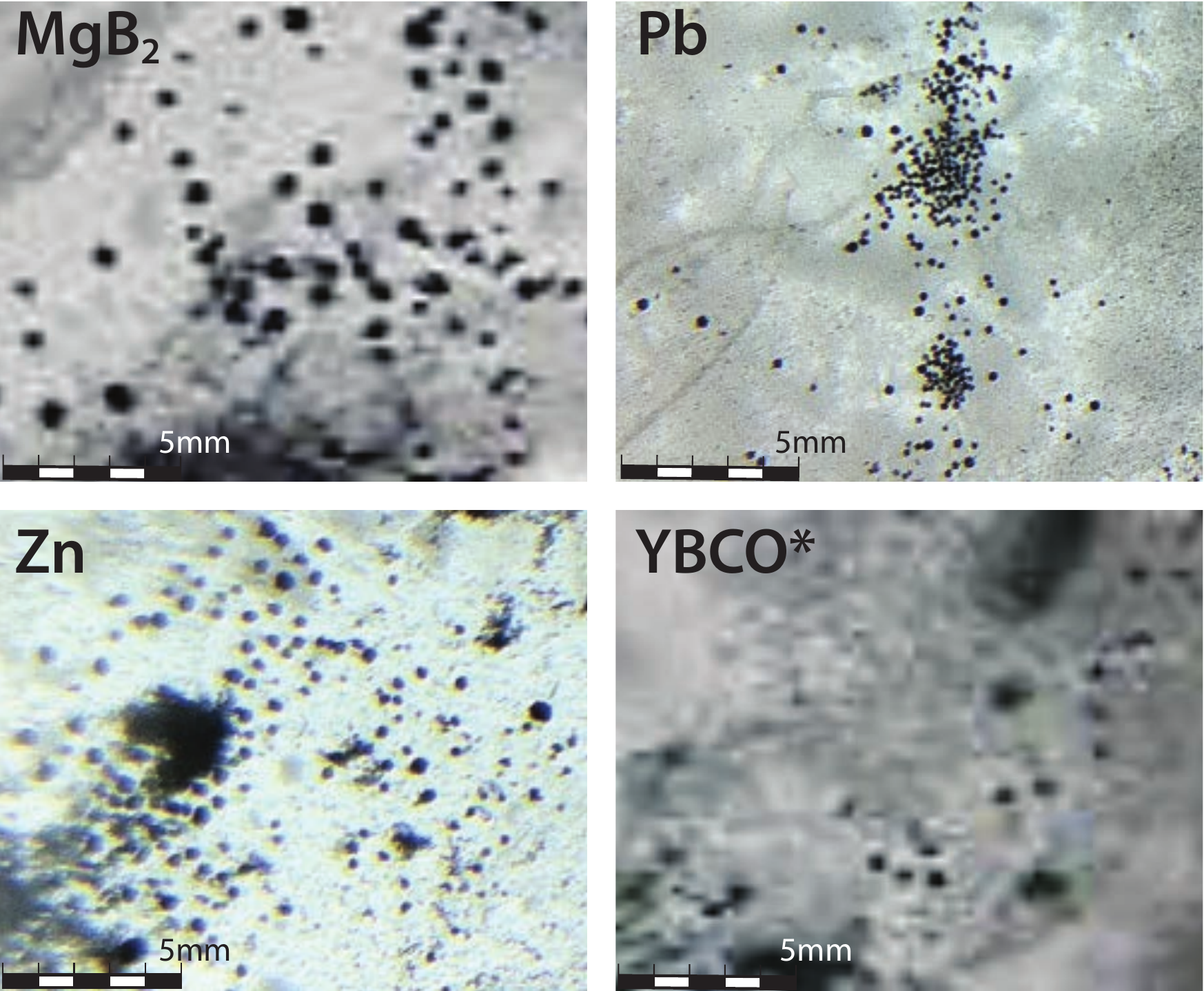}}
 \caption {Balls obtained by shaking powders in liquid nitrogen: $MgB_2$, $Pb$, $Zn$ and underdoped $YBCO$ (green phase).}
 \label{pbzn}
 \end{figure}

Next we turned to the conventional  temperature superconductors $MgB_2$ ($T_c=39K$), $Pb$ ($T_c=7.2K$) and $Zn$ ($T_c=0.875 K$), samples (3), (4)  and (5). It was found that balls still formed
under shaking, however
the agglomeration process took several more minutes and the resulting balls for $Pb$ and $Zn$ were considerably smaller than in the other cases, of diameter less than $0.3 mm$.
We show a snapshot of the balls formed in Fig. \ref{pbzn}. We also show in Fig. \ref{pbzn} balls obtained from insulating $YBCO$ (green phase) (see sect. IV.A. (2)).

\begin{figure}
 \resizebox{8.5cm}{!}{\includegraphics[width=9cm]{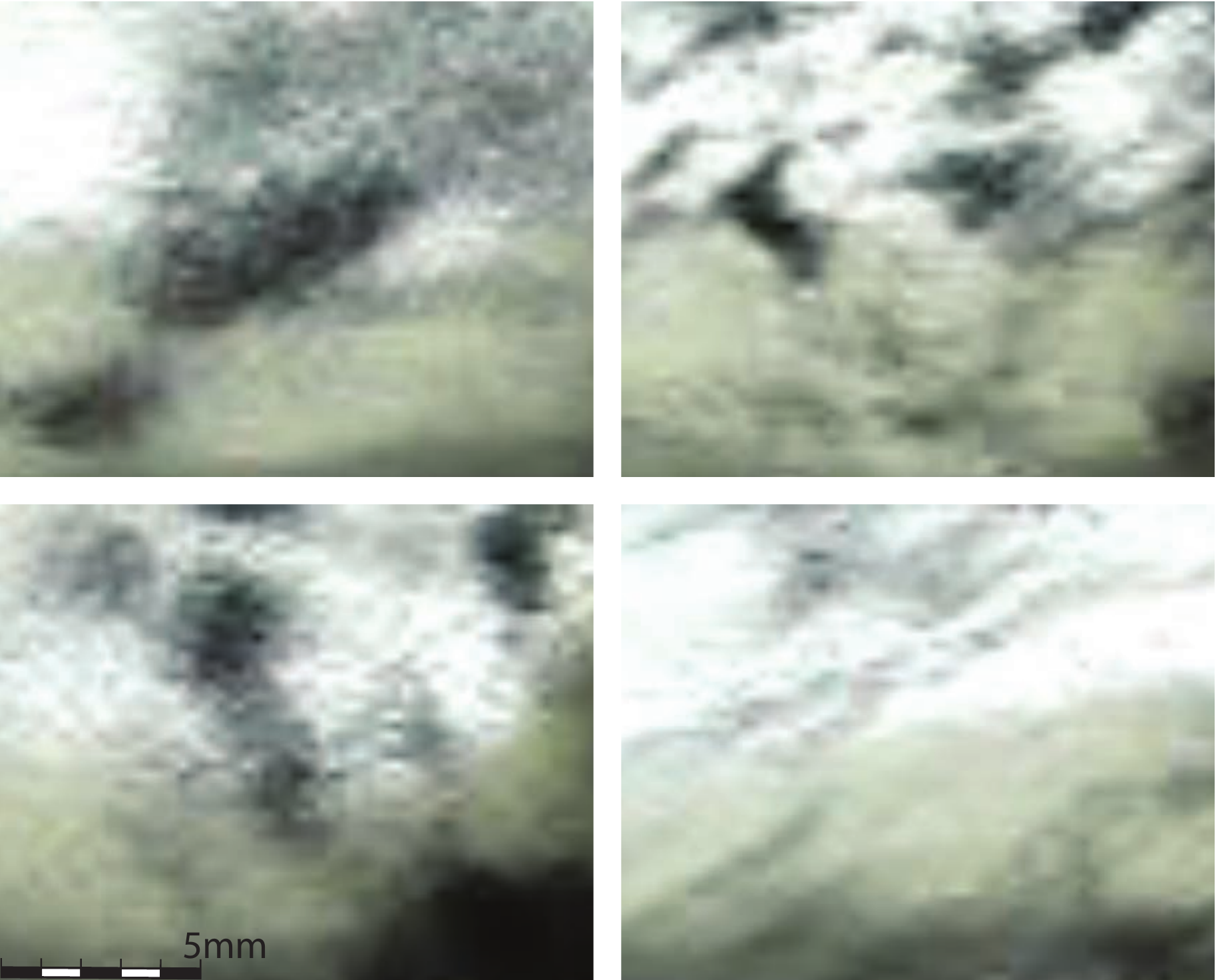}}
 \caption {Shaking of $BSCCO$ micrograins in water at room temperature does not result in agglomeration and ball formation, the powder remains highly dispersed.
 The four panels show snapshots separated by several minutes of shaking each.}
 \label{water}
 \end{figure}

 We also attempted to form balls by the same shaking procedure in  water at room temperature. No tendency to agglomeration was found despite repeated attempts.
 Figure  \ref{water} shows examples of snapshots obtained during the shaking process for $BSCCO$ micrograins (size less than $32 \mu m$ as always) obtained from grinding
 the $BSCCO$ disk with transition temperature above $100K$. No tendency to agglomeration whatsoever was found despite repeated attempts, as figure  \ref{water} illustrates.
 We repeated these experiments in ethanol and methanol close to their solidification temperature (159 K and 175 K) and found behavior similar to that shown in figure  \ref{water}, with no
 tendency to ball formation.
 
    \section{nature of the balls formed} 
    
    The balls formed by shaking or in the capacitor remain stable when the liquid evaporates and the container heats to room temperature, provided the heating proceeds
    slowly enough to prevent violent boiling. In this section we discuss some properties of the balls obtained for the case of high temperature superconductors ($BSCCO$ and $YBCO$).

       \begin{figure}
 \resizebox{8.5cm}{!}{\includegraphics[width=9cm]{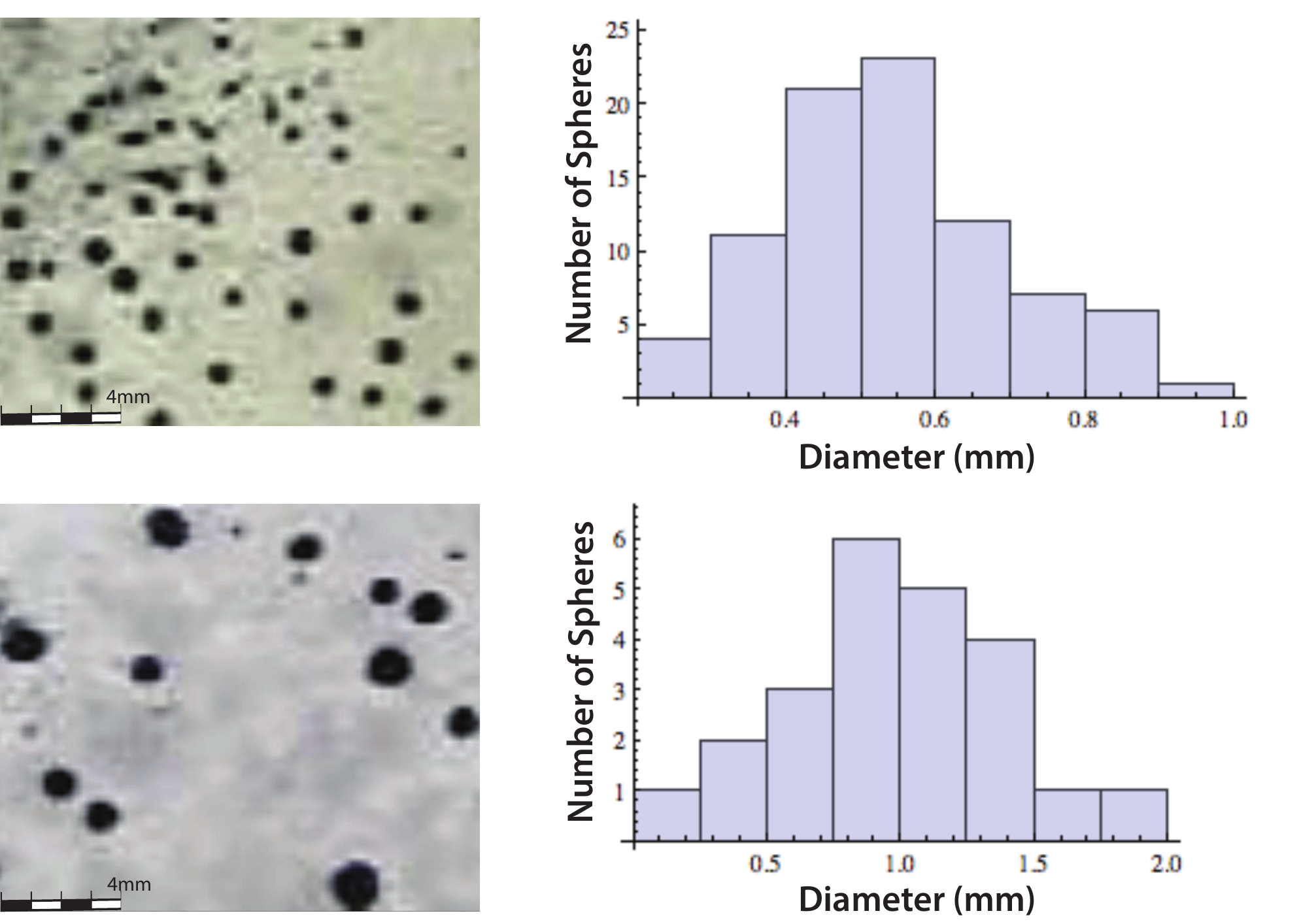}}
 \caption {Two examples of balls formed by shaking $BSCCO$ in nitrogen. The smaller balls (top figure) formed first after about two minutes of shaking, the bottom figure 
 showing bigger balls is a snapshot taken after approximately two more minutes of shaking. The right side of the figures shows histograms of the ball sizes for both cases.}
 \label{size}
 \end{figure}

\begin{figure}
 \resizebox{7.5cm}{!}{\includegraphics[width=9cm]{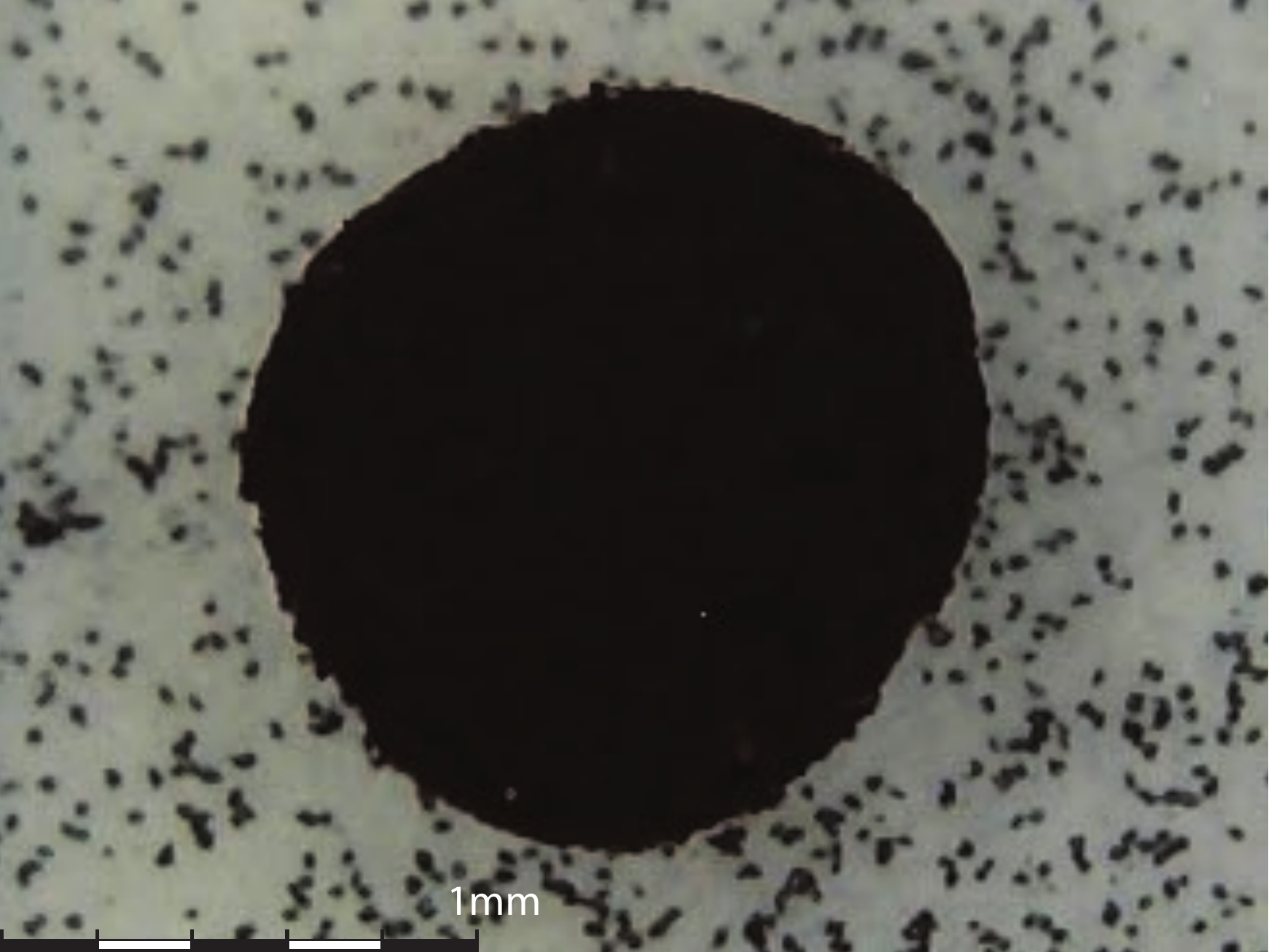}}
 \caption { A typical $BSCCO$ 
  large ball of diameter approximately $1.5 mm$ formed by microparticles smaller than $32 \mu$, surrounded by microparticles  that passed through a $32 \mu m$ sieve but did not pass through a $25 \mu m$ sieve. 
 The microparticles forming the ball are of this size and  smaller.}
 \label{largeball}
 \end{figure}

\begin{figure}
 \resizebox{7.5cm}{!}{\includegraphics[width=9cm]{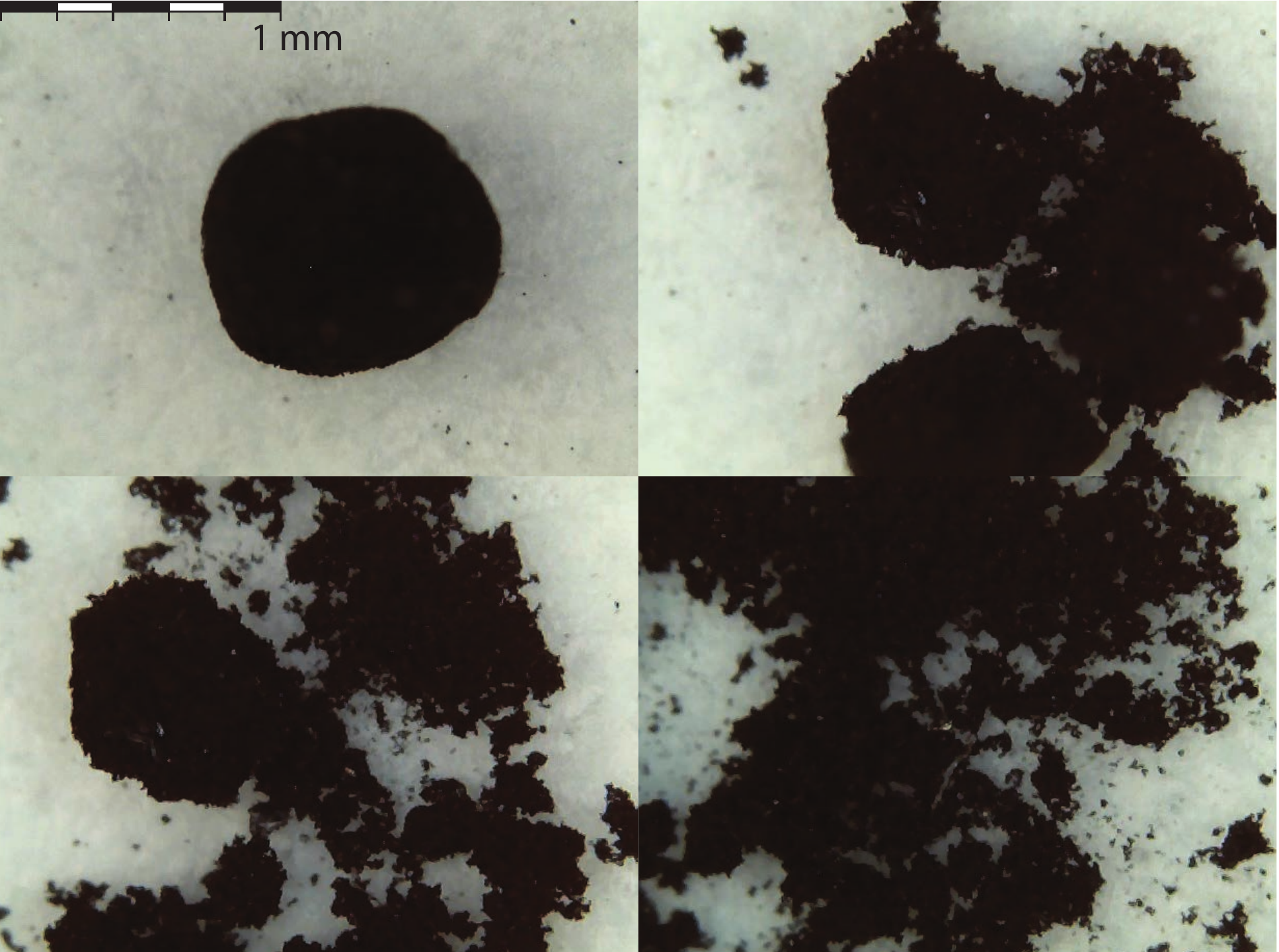}}
 \caption {A typical $BSCCO$ ball of diameter approximately $1 mm$ in varius stages of decomposition. }
 \label{decomp}
 \end{figure}

    In a typical run we obtain a distribution of ball sizes. Figure  \ref{size} shows an example. The maximum ball size found was of order 2 mm diameter, both in the cases where
    balls were formed by shaking as well as in the capacitor.
       Figure  \ref{largeball} shows a typical large ball, of diameter 1.5 mm, surrounded by microparticles of diameters between $25 \mu m$ and $32 \mu m$. 
    All microparticles used in forming this and other balls passed through a $32 \mu m$ sieve, as discussed in section II, hence none are larger than those shown in Fig. 10.
    However most of the microparticles forming the balls are substantially smaller than those shown in Fig. 10 surrounding the ball, as illustrated in Fig.  \ref{decomp} where we
 show a ball of 1 mm diameter in various stages of
    decomposition.
    
\begin{figure}
 \resizebox{7.5cm}{!}{\includegraphics[width=9cm]{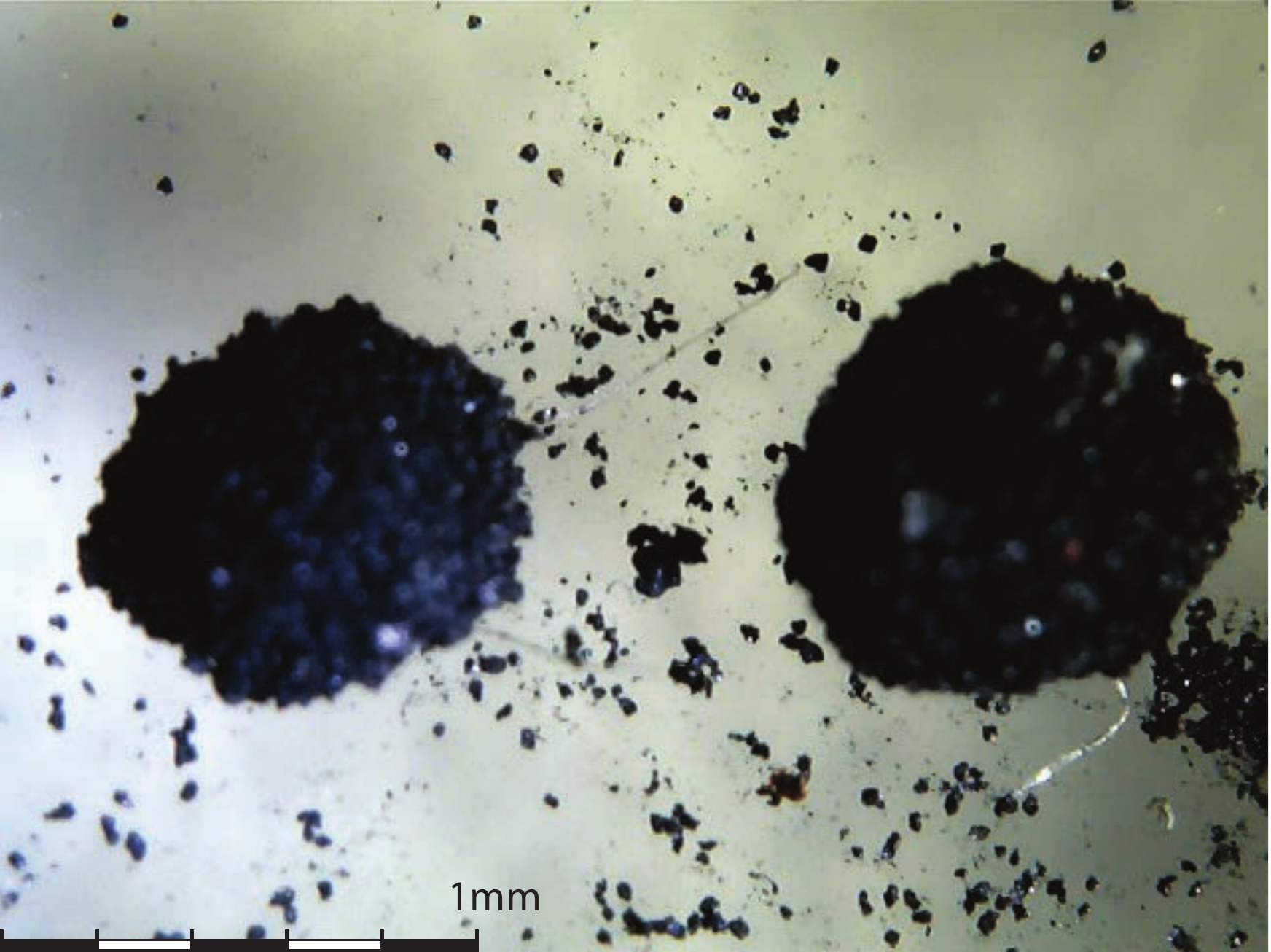}}
 \caption { Two balls formed from agglomeration of particles of sizes between $25 \mu m$ and $32 \mu m$ only. }
 \label{2balls}
 \end{figure}

    It appears that the presence of small microparticles is essential to the process of ball formation. We attempted to form balls composed of microparticles of large sizes only, and found
    that balls didn't form when the minimum size of the microparticles was $32 \mu m$ or larger. We did succeed after shaking gently for several minutes to form balls
    with microparticles restricted to the range $25 \mu m$ to $32 \mu m$. Figure  \ref{2balls} shows examples of such balls obtained. Note that the surface of a ball here is rougher
    than that of  the larger ball shown in figure  \ref{largeball}, where the powder contained also a large number of microparticles smaller than $25 \mu m$. In addition to being
    more difficult to form, the balls formed
    when the micrograin size was restricted to larger than $25 \mu m$ were found to be substantially more fragile than those that included smaller micrograins.

    We measured the density of the balls at room temperature for $BSCCO$ in various runs, and found that it ranged between  $1.2 g/cm^3$ and $1.6 g/cm^3$. 
    In contrast, the dry powder of microparticles of diameter $< 32\mu m$ used in forming the balls,
    when put in a glass tube and gently shaken, reached a  density  between  $1.8 g/cm^3$ and $2 g/cm^3$, and the disks from where the powders were obtained had density
    between $3.3 g/cm^3$ and $3.6 g/cm^3$. This indicates  that the balls are quite   loosely assembled and  have significant voids between micrograins. There was no apparent difference in the density of balls formed by shaking and in the capacitor.
    
Nevertheless, the balls are remarkably stable, both in liquid nitrogen and in air at room temperature.
    We found that at room temperature balls that formed either by shaking or in the capacitor  could be dropped from a height of    at least $10cm$ without breaking. The
    measured  terminal velocity was of order
     $80 cm/s$. In the capacitor, we found that balls bouncing back at forth at speeds of order $10 cm/s$ were stable, and in the shaking experiments balls moving at
    that speed were also stable.

 \section{non-superconducting materials}
 In the previous sections we have found that superconducting material microparticles aggregate into spheres in liquid nitrogen and argon
 both in the presence and in the absence of electric fields, as
 well as below and above their superconducting transition temperatures. The question then arises, can spherical aggregation also occur with nonsuperconducting particles?
 
\begin{figure}
 \resizebox{7.5cm}{!}{\includegraphics[width=9cm]{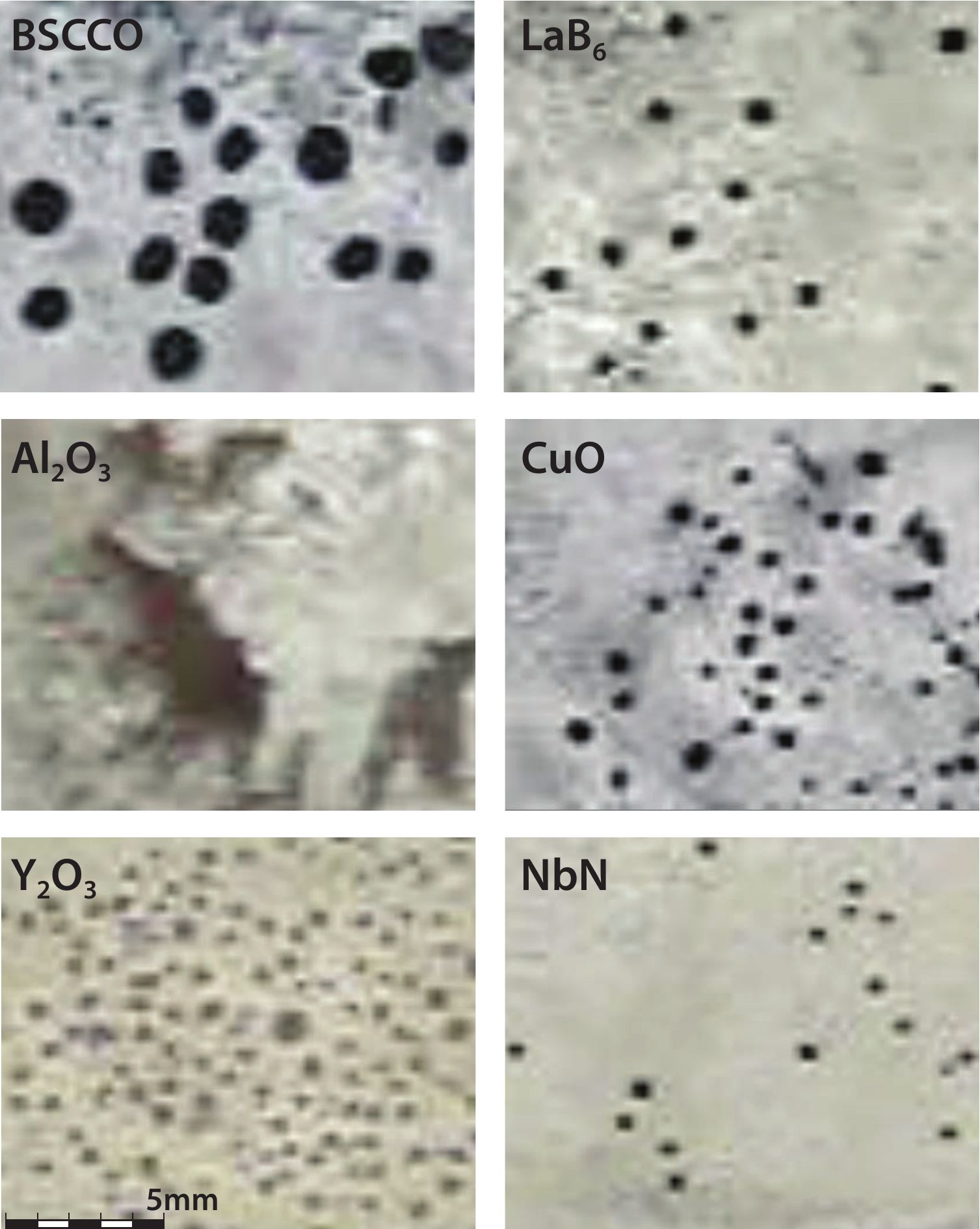}}
 \caption { Spherical aggregation of various substances  obtained by shaking of microparticles in liquid nitrogen. }
 \label{nonsc}
 \end{figure}

 The answer is yes. Figure  \ref{nonsc} shows results for 5 different substances, as well as for superconducting $BSCCO$, obtained 
 following the same  shaking procedure 
  described earlier. For all materials the size of microparticles used was considerably smaller than $32 \mu m$.  Of these materials we found that only for 
 $Al_2O_3$ there seemed to be no tendency for spherical aggregation. 
For  the other materials aggregation was found and the spheres obtained were of
 diameter up to $\sim 1 mm$.While it took different amounts of time to get aggregation,   the balls formed
 appeared to be very similar to those found in superconducting materials, as Figure \ref{nonsc}  illustrates.

    \section{the physics of spherical agglomeration} 
  
Spherical agglomeration of both superconducting and non-superconducting particles has been previously reported \cite{tao,stock}, however  the explanations for the agglomeration \cite{tao,farnand} are incapable of explaining the phenomenon presently shown. Prior explanation of the aggregation of superconducting microparticles argued that the presence of an external electric field created a new type of surface tension in superconducting particles which caused them to aggregate into spheres. However, we have shown in this paper
  that these same microparticles will aggregate even in the absence of an electric field, which discredits this argument. Agglomeration of non-superconducting microparticles has previously been attributed to bridging, which requires the use of two fluids as well as the microparticles. In the experiment here only $N_2$ (or $Ar$)  is present in liquid form and it has been shown that water from the atmosphere cannot act as a bridging agent significantly below it's melting point \cite{smith} and the nitrogen was kept from boiling so there was also no gas-liquid surface to facilitate bridging. Agglomeration on non-superconducting microparticles has been reported in the absence of a bridging agent for sub-micron particles \cite{ford}, and 
  attributed   to attractive Van Der Waals forces between the particles.  
The aggregation reported here of particles considerably larger  than sub-micron cannot be explained by these traditional methods indicating a new process that produces spherical agglomeration in a wide variety of microparticles at liquid nitrogen and argon temperatures.

            \begin{figure}
 \resizebox{8.5cm}{!}{\includegraphics[width=9cm]{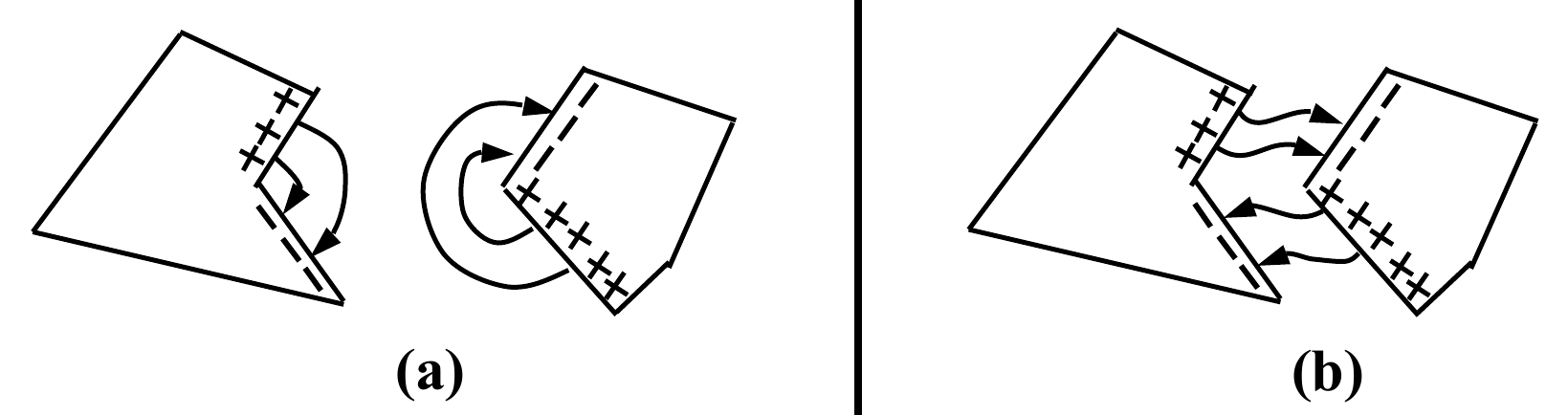}}
 \caption { Schematic depiction of charged planes of neighboring microparticles and surrounding electric field lines
 for (a) microparticles far apart and (b) microparticles closer together. }
 \label{forces}
 \end{figure}
 
 We conjecture that the forces giving rise to agglomeration in the cases considered here are predominantly electrostatic. The agglomeration appeared to be strongest
 for materials that are also high temperature superconductors. These materials (including $MgB_2$) have as a common feature the presence of $planes$ with a large
 charge imbalance. The high $T_c$ cuprate materials have negatively charged $CuO_2$ planes, with two excess electrons per unit cell, and compensating
 positively charged planes in-between. $MgB_2$ has negatively charged $B^-$ planes and positively charged $Mg^{++}$ planes. For an irregularly shaped
 microparticles there are likely to be sections where either a negatively or a positively charged plane is closest to the surface, and as a consequence
   electric field lines going from the negatively to the positively charged surface of a microparticle. It would then be energetically advantageous to join
 the electric field lines of neighboring microparticles as  in the arrangement shown in   figure \ref{forces}  and bring the microparticles close together.  
 This physics is related to the fact that quite generally a crystal will have different work functions for surfaces parallel to different crystal planes, and hence an electric
  potential difference
 between different surfaces and electric fields around them. We conjecture that the layered high temperature superconducting materials have larger differences between
 workfunctions for different crystal planes than other materials.

 As a rough estimate of the energetics involved, consider two surfaces of neighboring microparticles facing each other, of area $A=(10\mu m)^2$ at a distance
 (determined by the surface roughness) of $d=1\mu m$, with a potential difference of $5V$. The electrostatic energy associated with this is $U=CV^2/2\sim 10^{-14}J$, with
 $C=\epsilon_0A/d$. If these are grains of density $\rho=3g/cm^3$ and volume $V=(10\mu m)^3$ moving at speed $v=10cm/s$, their kinetic energy
 is $K=1/2\rho V v ^2\sim 10^{-14}J$, i.e. of the same order. This is consistent with the fact that we observe the balls to be stable in liquid nitrogen for speeds of that order and become unstable
 for larger speeds. For microparticles of larger volume and mass, the kinetic energy increases proportionally to the volume  while the potential energy gained in joining
 microparticles will increase not faster than the surface area, hence   the speed at which they break apart should become smaller, consistent
 with our observations.

   The hypothesis that the forces causing agglomeration are electrostatic is also consistent with the fact that we did not find agglomeration in water (figure \ref{water}).
 Water has negative $OH^-$ and positive $H^+$  (or $H_3O^+$) ions  in solution, and these ions should screen the electric fields near the surface of microparticles and
 inhibit the aggregation process. The microparticles themselves can also create further ionization of water in addition to the self-ionization so that no aggregation
 should occur even with a high concentration of microparticles. The same physics presumably inhibited aggregation in the solutions with  ethanol and methanol.
 
      \section{summary and conclusions} 
  
In this paper we reported results showing that: (i) in a capacitor, ball formation is induced by application of a high
voltage between the plates if the electrodes
are conducting but not if they are insulating, (ii) aggregation of microparticles into
balls  occurs also  in the absence of
an applied electric field, (iii) balls formed below $T_c$ remain robust when the temperature is
raised above $T_c$, (iv) balls form  both  below and above $T_c$ in liquid $N_2$ and liquid $Ar$ for a variety of cuprate superconductors,   (v) $MgB_2$, $Pb$, and $Zn$ agglomerate into spheres at temperatures well above their respective $T_c$, (vi) insulating $YBCO$ also agglomerates into spheres in liquid nitrogen,
(vii) microparticles of several  different non-superconducting materials suspended in liquid nitrogen also aggregate into spheres of $mm$ dimensions upon shaking.

Our findings show significant differences with the findings reported by Tao et al\cite{tao,tao2,tao3,tao4}. Tao et al  reported that
(1) an electric field drives formation of superconducting  balls,  (2) the balls become fragile if the electric field is turned off, (3) there is a lower critical electric field $E_{c1}$, that is frequency dependent, 
below which balls don't form,  (4) the radius of balls decreases when the electric field increases, 
(5) in a large electric field, powder that clings to the electrodes below $T_c$  will fly off above $T_c$, 
(6) balls of $BSCCO$ formed in nitrogen break apart when liquid argon is added to    increase   the temperature, (7) balls don't form
 above the superconducting transition temperatures,  (8) balls formed at liquid nitrogen temperature become fragile at room
temperature, (9) granular particles making the balls are closely packed, (10) the surface of the grains needs to be conducting for balls to form. Each one   of 
these findings is  in 
  disagreement with our findings. 

Additionally, Tao et al   reported that the radius of the balls is a function of the AC frequency of the electric field,   that the radius of $MgB_2$ balls goes linearly to zero as the temperature
approaches $T_c$ from below, and that balls break up when a strong magnetic field is applied. We have not tested these findings directly but conclude in view of our other findings that they are unlikely to be correct.

In their study of ball formation in low frequency ac electric fields\cite{tao3}, Tao et al reported that the lower critical field for ball formation, $E_{c1}$, decreased strongly
as the frequency decreased, and jumped almost discontinuously  to a high value  at zero frequency (see Figs. 6, 7  and 8 of Ref. \cite{tao3}). For example,
for $NdBa_2Cu_3O_x$, $E_{c1}$ decreased smoothly from about $1200 V/mm$ for frequency $500 Hz$ to $117 V/mm$ for $10Hz$, and jumped to $760 V/mm$ for a
dc field.  Our interpretation of these observations is that the ac field causes ``shaking'' of the powder in a way similar to our shaking experiments. For higher frequency
a larger electric field is needed so that the displacement of the microparticles is sufficient to cause agglomeration, and for very low frequencies the amplitude of the
electric field required approaches zero, consistent with our observation that agglomeration occurs also in the absence of electric fields.

Our findings indicate that spherical aggregation of microparticles (both conducting and insulating) suspended in liquid nitrogen and liquid argon is a very general phenomenon.
We did not find clear evidence that ball formation   is helped
by the presence of  superconductivity in the sample as predicted theoretically\cite{taoeffect}. We did generally find that the easiest and fastest ball formation, particularly
of large balls, occurred for microparticles of high $T_c$ superconductors ($BSCCO$ and $YBCO$).
 We hypothesized that a strong  tendency to ball formation results from the existence of charged planes in the structures.
Whether or not there are subtle changes in the tendency to ball formation 
depending on whether the system is in the superconducting or normal state  remains an open question.

  \acknowledgements{}
  The authors are grateful to B. Maple and his group members, particularly Ivy Lum and Colin McElroy, 
   for kindly sharing laboratory supplies, pressing $\sigma-$powder into pellets,  and  measuring the resistive transition temperature of the 
annealed $\sigma$ sample,   to Ali Basaran for assistance, and to the UCSD Physics Department for financial support.

\end{document}